# An Alternative View of the Universe Structure
## (on the invalidity of the four dimensional space-time concept)

Felix Hovsepian

The model of Universe uses equations of the unperturbed Keplerian motion of celestial bodies generalized, updated and elaborated using the notion of the characteristic function of a random value from the theory of probabilities. The differential equation argument in such a model will be not time t, but an interval $\tau$ between sections of the stationary random process at arbitrary times $t_1$ and $t_2$. It is proved that $\tau$ is not a function of space which immediately shows the invalidity of the concept of the four dimensional space-time. As a result the picture of the Universe structure changes radically: it becomes Euclidean and stationary. It is also proved that $\tau$ possesses a number of other non-trivial features.

The author cites astronomical observations which support mathematical analysis of the Universe model elaborated in the present paper.

## 1. Introduction

The model of the Universe in this paper uses equations of the unperturbed Keplerian motion. These equations were obtained by the great I. Newton after he formulated the law of gravitation, derived the Keplerian laws of planets motion and then translated them to the language of differential equations. It became later known that stars in star constellations, satellites, comets, etc obey these equations.

Theoretical astronomy founded by I. Newton has been for almost 250 years the model of an exact science. It had reached what was a dream of the mankind since times immemorial. It had raised the curtain over the future endowing its disciples with the gift of foresight. However, the laws of the Universe do not allow exceptions. Still, there is one instance where the theory failed. Even though the mistake was small it was irrefutable. It was the case with the planet Mercury.

The above mistake was one of the main reasons for substitution of two postulates by I. Newton with two other postulates by A. Einstein which resulted in the four dimensional space-time.

Even though equations of the unperturbed Keplerian motion are well-known in the celestial mechanics, they have been generalized, updated and elaborated further using the notion of the characteristic function of a random value [1].

*Definition* 1. Let $\Omega$ be a random value with probability distribution $F(\omega)$. Then the characteristic function of distribution $F(\omega)$ (or random value $\Omega$) is called function $h(\tau)$, determined for real-valued $\tau$ from the formula

$$h(\tau) = \int_{-\infty}^{\infty} e^{i\tau\omega} dF(\omega) = r(\tau) + id(\tau).$$

In what will follow we will consider a particular case of this definition, when function F(ω) is differentiable, i.e. the random value Ω has probability distribution V(ω), whereas function h(τ) is an even function. Hence

$$h(\tau) = 2\int_{0}^{\infty} V(\omega)\cos\omega\tau d\omega. \quad (-\infty < \tau < \infty) \tag{1.1}$$

Characteristic functions have been known in the theory of probabilities for a long time. In the early 1930s these functions attracted once again the attention of scientists who studied stationary random processes. The necessary and sufficient conditions have been determined for an arbitrary function to be presented as (1.1) [1].

Characteristic functions have started being applied to solution of deterministic problems quite recently. The approach the present author uses for obtaining a characteristic function implies that the solution y(t) of a linear asymptotically stable differential equation of second order is given in the form of Fourier transformation. To calculate such an integral one needs a function in the range (−∞, ∞), hence, function h(τ) is determined which consists of y(t) for t > 0 and y(t) extrapolated evenly to t < 0. For certain initial conditions function h(τ) assumes the form (1.1) [2 - 4].

Solutions h(τ) in the form of (1.1) enable the author to develop the model of Universe given in the paper. Given mathematical results, physical interpretation of some of them, as well as astrophysical observations cited in the paper, one can state that characteristic functions treated as a solution of a differential equation is a finding of potentially great implications for theoretical physics and astrophysics in future.

Numerous astronomical supervision and the researches lead by known astrophysicist
N.A.Kozyrev
personally or under his management, confirm adequacy of the model in the present paper [5 – 8].

Any text inserted into the theoretical part of the paper is marked with the sign ■.

## 2. Informal statement of the problem

Consider equations of the unperturbed Keplerian motion which introduce the problem of the motion of a material point with mass $m_0$ due to the Newtonian gravity of a centrally located body with mass $M_0$, which is also treated as a material point:

$$\frac{d^2x}{dt^2} = -\frac{\mu x}{r^3}, \qquad \frac{d^2y}{dt^2} = -\frac{\mu y}{r^3}, \qquad \frac{d^2z}{dt^2} = -\frac{\mu z}{r^3}, \tag{2.1}$$

where μ is a constant value which is a function of $M_0$ and $m_0$, whereas

$$r = \sqrt{x^2 + y^2 + z^2}$$

is a radius-vector of a moving point. The set of equations (2.1) is sometimes more conveniently referred to as the system ($M_0 - m_0$).

From the point of view of the theory of probabilities and mathematical statistics the author found in (2.1) «mistakes», which are as follows. Any of the equations (2.1), for instance, along the y coordinate, is a particular form of the second-order equation

$$\frac{d^2y}{dt^2} + A\frac{dy}{dt} + By = 0, \qquad (2.2)$$

where $A = 0$, and $B > 0$. The characteristic equation (2.2) can have three types of roots:

a) one real-valued double root;

b) two different real-valued roots;

c) a pair of complex-conjugate roots.

In cases a) and b), assuming $A = 0$, we obtain either $B = 0$, or $B < 0$, i.e. <u>it is not feasible</u> to obtain equations of motion (2.1). These equations can be obtained only in case c), i.e. in the case of only one pair out of three possible. <u>The probability of the event considered equals one third</u>.

Why this particular pair of roots was given by the Nature the preferential right to represent the unperturbed Keplerian motion? One cannot see any objective reason for prevailing of this particular pair in the equation (2.1), which means that *<u>the laws of the probabilities theory have been violated.</u>*

Now assuming that the period of rotation of the Galaxy in the region of the Sun equals approximately 275 million years, and it is essentially "a pancake" with the following approximate sizes: thickness 2500 parsec, diameter 30000 parsec and the Milky Way alone has billions moving celestial bodies, whose trajectories are described by the solution of equations (2.1), it becomes obvious that *<u>the laws of mathematical statistics have also been violated significantly.</u>* Both, equations of the unperturbed Keplerian motion and laws of the theory of probabilities and mathematical statistics have been copied from Her Majesty Nature, and the Nature cannot be wrong. It immediately follows that Nature is not to be blamed for "the missing two pairs of roots". Hence, equations (2.1) should exist in a different form and one has to find it. This is the statement of the goal of the investigation.

It is almost obvious that the solution for the "not yet found equations" of type (2.1) should be the even function, which is why function $h(\tau)$ already discussed in the introduction was obtained.

### 3. Invalidity of the four dimensional space-time concept

According to theorem (see Appendix 1) a solution of the second order differential equation for certain initial conditions can be presented as the Fourier integral (1.1):

$$h(\tau) = 2\int_0^\infty V(\omega)\cos\omega\tau\, d\omega, \quad (-\infty < \tau < \infty)$$

where functions $h(\tau)$ and $V(\omega)$ have indices which depend on the type of the characteristic equation's roots. For further discussion the specific type of these functions is not relevant, hence, we will skip them.

From theorem 1 we know that $v(0) = 1$, hence

$$h(0) = 1 = 2\int_0^\infty V(\omega)d\omega = \int_{-\infty}^\infty V(\omega)d\omega,$$

consequently, $V(\omega)$ is a normalized density of the probability distribution of a random value $\Omega$ with mathematical expectation

$$E\Omega = \int_{-\infty}^\infty \omega V(\omega)d\omega = 0, \tag{3.1}$$

because the integrand function is odd.

Let us differentiate function $h(\tau)$ twice with respect to $\tau$ under the sign of integration as a parameter (this can be done by virtue of theorem 2 from Appendix 2) and, substituting in the second-order equation

$$\frac{d^2h}{d\tau^2} + a\frac{dh}{d\tau} + bh = 0, \quad (\tau > 0) \tag{3.2}$$

whose characteristic equation can be written as

$$\eta^2 + a\eta + b = 0,$$

we obtain identity

$$-\int_0^\infty \omega^2 V(\omega)\cos\omega\tau\, d\omega - a\int_0^\infty \omega V(\omega)\sin\omega\tau\, d\omega + b\int_0^\infty V(\omega)\cos\omega\tau\, d\omega \equiv 0. \tag{3.3}$$

The model of Universe elaborated in this paper is the differential equation (3.2) with the solution (1.1), i.e. the model is practically the identity (3.3). Assuming that $\tau = 0$ in (3.3) and extending the limits of integration, we obtain:

$$\int_{-\infty}^\infty \omega^2 V(\omega)d\omega = b\int_{-\infty}^\infty V(\omega)d\omega = b,$$

hence

$$D\Omega = \int_{-\infty}^\infty \omega^2 V(\omega)d\omega = b, \tag{3.4}$$

where $D\Omega$ is the dispersion of random value $\Omega$.

Extremely important for applications of (1.1) are two limit theorems – direct and inverse [1]. By virtue of these theorems the correspondence in (1.1) between functions $V(\omega)$ and $h(\tau)$, is one-to-one. To study this relation we substitute the integration variable

$$\omega = \frac{\omega_1}{\alpha} \quad (\alpha > 0) \tag{3.5}$$

and write $V(\omega)$ as

$$V(\omega) = V(\frac{\omega_1}{\alpha}) = \overline{V}(\omega_1).$$

Before making substitutions in the identity (3.3), let us clarify the physical meaning of the substitution (3.5), because this is going to be <u>extremely important</u> for what will follow. The abscissa axis in the coordinate system $(\omega, V(\omega))$ expands for $\alpha > 1$ or contracts for $\alpha < 1$. Since the area under the distribution curve does not change and remains equal to unity, then $V(\omega)$ should change becoming either more flat or more acute depending on the specific value of $\alpha$. This way we obtain a new density $\overline{V}(\omega_1)$ with a different dispersion. True, substitute (3.5) in the identity (3.3):

$$-\int_0^\infty \omega_1^2 \overline{V}(\omega_1) \cos(\frac{\omega_1}{\alpha}\tau) d\omega_1 - \alpha a \int_0^\infty \omega_1 \overline{V}(\omega_1) \sin(\frac{\omega_1}{\alpha}\tau) d\omega_1 + \alpha^2 b \int_0^\infty \overline{V}(\omega_1) \cos(\frac{\omega_1}{\alpha}\tau) d\omega_1 \equiv 0. \tag{3.6}$$

Assuming that $\tau = 0$ in (3.6) and extending the limits of integration, we obtain an equality

$$D\Omega_1 = \int_{-\infty}^\infty \omega_1^2 \overline{V}(\omega_1) d\omega_1 = \alpha^2 b \int_{-\infty}^\infty \overline{V}(\omega_1) d\omega_1 = \alpha^2 b, \tag{3.7}$$

i.e. <u>from the point of view of the probabilities theory</u> the substitution (3.5) resulted in the change of the dispersion (3.5) up to the value of (3.7).

The substitution (3.5) has caused changes <u>from the point of view of the differential equations</u> too: the coefficient b in the equation (3.2) has changed into $\alpha^2 b$. The change in the coefficients of (3.2) leads to a change of its characteristic equation, i.e. a change in the roots of that equation. To find out the real change the roots have undergone, we will assume that prior to the substitution in the characteristic equation we had complex-conjugated roots $-\chi \pm i\lambda$. It follows from the Vieta's theorem that the roots product should be equal to the coefficient b:

$$\chi^2 + \lambda^2 = b.$$

Denote the root values after the substitution as $-\chi_1 \pm i\lambda_1$, then the product of the new roots by virtue of the same theorem is

$$\chi_1^2 + \lambda_1^2 = \alpha^2 b.$$

Comparing the results we obtain:

$$\chi_1 = \alpha\chi, \qquad \lambda_1 = \alpha\lambda.$$

The change in the roots should lead not only to the change in the coefficient b in equation (3.2), but also to a change in the coefficient a, which is the case. This can be easily seen assuming that in identity (3.6) $\tau = \alpha\tau_1$ and reduce it to the following form

$$-\int_0^\infty \omega_1^2 \overline{V}(\omega_1) \cos(\omega_1\tau_1) d\omega_1 - \alpha a \int_0^\infty \omega_1 \overline{V}(\omega_1) \sin(\omega_1\tau_1) d\omega_1 + \alpha^2 b \int_0^\infty \overline{V}(\omega_1) \cos(\omega_1\tau_1) d\omega_1 \equiv 0. \tag{3.8}$$

Now note that we will obtain the same identity by substituting the Fourier integral

$$\overline{v}(\tau_1) = \int_0^\infty \overline{V}(\omega_1)\cos(\omega_1\tau_1)d\omega_1 \qquad (3.9)$$

in the second order differential equation of type (3.2), whose characteristic function, however, will have the following form

$$\eta^2 + \alpha a\eta + \alpha^2 b = 0.$$

Hence, the substitution (3.5) is simultaneously change of roots of the characteristic equation in (3.2) as a result of which vary such important from the point of view of management of the characteristic such as:

a) duration of the transition period;

b) over regulation;

c) the degree of stability, etc.

Experts in control systems develop special feedback tools to improve and update the above indicators. This means that a change in the characteristic equation as a result of the substitution of (3.5) cannot <u>in principle</u> cause the above changes. Since such a change has really happened then automatically the argument $\tau$ in $\tau_1$ should automatically change:

$$\frac{\tau}{\alpha} = \tau_1, \qquad (3.10)$$

to make up for the change as it is shown:

$$\frac{d^2h}{d\tau^2} + a\frac{dh}{d\tau} + bh = 0 \;\;\rightarrow\;\; \frac{d^2h}{d\left(\frac{\tau^2}{\alpha^2}\right)} + \alpha a\frac{dh}{d\left(\frac{\tau}{\alpha}\right)} + \alpha^2 bh = 0.$$

Note also that the substitution (3.5) can be done anywhere in the Universe, i.e. this substitution does not depend on the space in any way. Consequently, the result of the substitution, i.e. (3.10) does not depend on the space either. It immediately follows from the result obtained that

*the concept of the four-dimensional space-time is not valid.*

Time t flows always in regular intervals, and $\tau$ depending on $\alpha$ can either be accelerated, or be slowed down, hence

*acceleration or deceleration of $\tau$ is not a function of gravity.*

Argument $\tau$, as will be shown in what follows, possesses a number of other nontrivial properties, Therefore, it seems reasonable to introduce a special name for it.

*Definition* 2. We will refer to argument $\tau$ as <u>flexible (elastic) time</u>.

Note that a moment of flexible (elastic) time $\tau$ corresponds to a point in space. As a result of substitution (3.5) this point in space <u>*instantaneously*</u> moves to another point of the same space where flexible time equals $\tau_1$. Obviously this transformation cannot be related to the translocation of the material body since τ and $τ_1$ can be separated by a space of arbitrary length (parameter α in (3.10) is arbitrary). This means that, on the one hand, the translocation in space has been done with an infinitely high speed. On the other hand, this means that we deal here with *a field of unknown origin*, *furthermore, it seems to be impossible to mask it.*

■*Astronomical observations*. Note that the word «time», mentioned in this and other observations is identified here with the notion of flexible (elastic) time introduced by the present author. The results obtained are supported with the astronomic observations given below.

[6], p. 87: «… long-range interaction is possible through time, i.e. *instantaneous communication* (marked out by the author). This conclusion was proved by observations of a resistor in the focal plane of the telescope responding to the region of the sky where no star is seen, however, where it is present at the moment of the observation. This star position can be easily calculated …»

[7], pp. 177-178: «During observations of Venus an unexpected phenomenon drew our attention: the measurement system registered not only the true position but also the image of Venus as seen by the telescope sight …Observations showed that when completely covering the index mirror with a duralumin screen …the intensity of the visible image was attenuated to the same extent as that of the true image … Consequently, the effect of the visible image *is not related to light* (marked out by the author), but only coincided with its direction». ■

Note that an old argument τ in (3.3) has changed to a new argument $τ_1$ in (3.10), but in such a way that the product ωτ remained unchanged:

$$ωτ = \left(\frac{ω_1}{α}\right)(ατ_1) = ω_1τ_1 = \text{const}. \qquad (3.11)$$

Totally unexpected one-to-one relation between the density V(ω) and the function v(τ): the frequency ω in the Fourier integral (1.1) and flexible (elastic) time τ are related between each other in such a way that the expansion ω leads to the contraction τ and, vice versa, the contraction ω results in the expansion τ, their product remaining unchanged.

In (3.11) values of ω and τ obviously can be chosen arbitrarily, however, having done so, we remain adhered to the chosen constant. Let us keep this result in mind, in para. 7 it will be used for theoretical proving of the fact that velocity of harmonic oscillation cosωτ is not constant.

■ *Physical interpretation.*

After substitution (3.5 )the identity (3.3) changes to (3.8), where the flexible (elastic) time acquires the form of (3.10). Two natural questions ensue.

   I. What is the substitution (3.5)?

   II. What is the final effect of the flexible (elastic) time in (3.10)?

At the answer to the first question we shall consider, that the dispersion and energy depend from each other. In more detail this question will be discussed in п.5.

Substitution (3.5) is a change in the energy state of the system ($M_0-m_0$). True, before the substitution this state was specified by dispersion (3.5). As a result of the substitution the dispersion has changed to (3.7). If $\alpha > 1$ in (3.7), then the amount of energy in the system ($M_0-m_0$) for $\tau = 0$ grows with the growing dispersion. However, if $\alpha < 1$, then the amount of energy in the system for $\tau = 0$ decreases with the decreasing dispersion.

The answer to the second question consists of three steps.

I. In resulted below the scheme the parameter $\alpha$ can be reduced, thus the differential equation acquires the previous form:

$$\frac{d^2h}{d\left(\frac{\tau^2}{\alpha^2}\right)} + \alpha a \frac{dh}{d\left(\frac{\tau}{\alpha}\right)} + \alpha^2 bh = 0 \quad \rightarrow \quad \frac{d^2h}{d\tau^2} + a\frac{dh}{d\tau} + bh = 0,$$

i.e. we will have $\tau$ instead of $\tau_1$.

II. We will have $\omega\tau$ instead of $\omega_1\tau_1$ by virtue of (3.11), i.e. $\omega_1$ changes to $\omega$.

III. Identity (3.8) changes to identity (3.3), as shown below:

$$-\int_0^\infty \omega_1^2 \overline{V}(\omega_1)\cos(\omega_1\tau_1)d\omega_1 - \alpha a \int_0^\infty \omega_1 \overline{V}(\omega_1)\sin(\omega_1\tau_1)d\omega_1 + \alpha^2 b \int_0^\infty \overline{V}(\omega_1)\cos(\omega_1\tau_1)d\omega_1 \equiv 0 \quad \rightarrow$$

$$-\int_0^\infty \omega^2 V(\omega)\cos(\omega\tau)d\omega - a\int_0^\infty \omega V(\omega)\sin(\omega\tau)d\omega + b\int_0^\infty V(\omega)\cos(\omega\tau)d\omega \equiv 0.$$

It means that $\tau$ adopting the form of (3.10) eliminates the consequences of the substitution done.

Consequently, the flexible (elastic) time possesses a physical property of removing the excess of energy in the system when it grows, and replenishing it when it decreases. Assume that there is an event taking place in the Solar system caused by a change in the state of energy on a planet. The information about this happening will immediately be transmitted to the Sun – this is the first system ($M_0 - m_0$). Sun in its turn will immediately pass over this information to the center of Galaxy – this is the second system ($M_0 - m_0$) and so on, cascading. Hence,

*Universe is a closed loop information-energy system which revives the nature and acts as a barrier to the second law of thermodynamics.*

If we learn how to control the flexible (elastic) time, i.e. if instead of $\tau$ we can obtain in some way (3.10) for $\alpha > 1$, then one can easily prove that this will cascade further on an increase of energy in the system ($M_0 - m_0$). This way we can obtain any amount of energy and, quite possible, that it will not be prohibitively costly. ■

■*Astronomical observations.*

[6], p. 86: «The fact that time possesses physical properties has been proved by a number of laboratory experiments and astronomical observations…Time carries certain organization or negative entropy which can be conveyed to another substance…»

[6], p. 93: «…active participation of time should revive the world and prevent its thermal death».■

■*N.A.Kozyrev's theoretical researches.*

In [5] in the chapter dealing with «Causal mechanics and a possibility of the experimental study of the time properties» on page 311, N.A. Kozyrev writes in his conclusions: «2. Lapse of time may create additional tension in the system and thereby modify its potential and total energy. This is why the lapse of time can be a source of energy. For astrophysicists this conclusion is very important since it may potentially be the underlying mechanism of starlight».

[5], p. 405: «Stars are omnipresent in the Universe. Hence, the explanation of their vitality should have the same affinity as only space and time have. However, no such possibility can be traced to the properties of space since the space is a passive arena for the events of the Universe to be enacted. It remains to conclude that time along with its passive geometric property measured in hours also possesses active physical properties thanks to which the time can interact with material systems and prevent their transition to an equilibrium state. So, time turns out to be a phenomenon of Nature…»

[5], p.384: «Time is a natural phenomenon with various features which can be studied by experimentation in laboratory and astronomical observations».

[5], p.286: «The Newton-Einstein mechanics and the nuclear mechanics result in the first and second laws of thermodynamics. Consequently, in the worlds matching these mechanics, only such processes are possible which promote the growing entropies eventually leading to thermal death. The real world, however … has peculiar features. This world can withstand the death through opposite processes which can be referred to as processes of life in the broadest sense of the word».

N.A.Kozyrev's researches confirm the results received on model ■

## 4. Correlation equations

The experts in differential equations are used to argument $\tau$ in (3.2) being time. In the case under consideration, however, this is not exactly true. To prove our conclusion we consider (3.1) and recall the well-known Khinchin's theorem [9]. It says: for the function $h(\tau)$ to be correlation function of a continuous stationary random process it is necessary and sufficient to be able to present it as (3.1), where $V(\omega)$ is the probability distribution density. $V(\omega)$ meets this requirement, which means that $h(\tau)$ is the correlation function of a random stationary process $\xi(t)$. In its turn it means that $\tau$ is the time interval between sections of $\xi(t)$ at arbitrary times $t_1$ and $t_2$:

$$\tau = t_1 - t_2. \tag{4.1}$$

i.e. $\tau$ can be treated as time but this is not quite an ordinary time.

■*About flexible (elastic) time.* At calculation $h(\tau)$ one of times in (4.1), we shall admit $t_1$, we shall accept for current time which will flow in regular intervals. And other time $t_2$ thus will be at the disposal of the researcher, i.e. it can be moved on an axis t forward or back with any speed. Time $t_2$ in relation to $t_1$ can accept both positive values, and negative. Thus flexible (elastic) time t in (4.1) can be equal to any certain value (for example, to zero: $t_1 = t_2$) during some time or will vary in regular intervals or non-uniformly up to as much as greater sizes, as all process - a handwork of the researcher. It means, that the Nature also can dispose t at own discretion. ∎

Integrand function in (3.9) on any frequency represents even function so the differential equation can be written as

$$\frac{d^2h}{d\tau_1^2} + \alpha^2 bh = 0, \qquad (4.2)$$

which is a correlation equation since its solution

$$\overline{\overline{h}}(\tau_1) = \overline{V}(\omega_1)\cos\omega_1\tau_1$$

is a correlation function where

$$\tau_1 = \frac{\tau}{\alpha}, \qquad \omega_1^2 = \alpha^2 b.$$

Let's prove, that $\overline{\overline{h}}(\tau_1)$ there is a correlation function. With this purpose we shall consider the stationary random process

$$\delta(t) = \alpha\cos\omega t + \beta\sin\omega t,$$

where α and β are random values with the following characteristics

$$E\alpha = E\beta = 0, \; E\alpha^2 = E\beta^2 = 1, \; E\alpha\beta = 0.$$

The correlation function of process δ(t) is

$$K(\tau) = E[(\alpha\cos\omega t_1 + \beta\sin\omega t_1)(\alpha\cos\omega t_2 + \beta\sin\omega t_2)] = \cos\omega t_1 \cos\omega t_2 + \sin\omega t_1 \sin\omega t_2 =$$
$$= \cos\omega(t_1 - t_2) = \cos\omega\tau = \cos\omega_1\tau_1,$$

if the relationship (3.11) is used.

Note that in equation (4.2):

a. if $\alpha = 1$, and $\tau_1 = t$, then we have one of the motion equations (2.1);

b. for arbitrary α and argument τ – this is the wave equation.

Hence, we can ascertain, that «two missing pairs of roots», discussed in para. 2 of this paper have been "newfound": the unperturbed Keplerian motion in the Universe described by equation (4.2), takes place for any type of roots of the characteristic equation in (3.2) (the probability of this event equals one).

On the basis of the received results we can ascertain:

*the Universe is stationary and, consequently, Euclidean*;

*time in the Universe - flexible (elastic) time τ.*

## 5. Source of energy for a celestial body

As it was already marked in the previous paragraph, the equation (4.2) is the equation of type (2.1), i.e. the equation unperturbed Keplerian motion. But (4.2) is the equation more the general type, first, than (2.1) as it takes place at any type of roots of the characteristic equation in (3.2). Secondly, the coefficient in (2.1) as it was marked in para. 2, depends on mass $M_0$ and $m_0$ moving celestial bodies, hence, in (4.2) coefficient b also depends on the same masses. Thirdly, according to (3.4) and (3.7) dispersion $D\Omega$ is proporsional to coefficient b in (4.2). Hence, dispersion $D\Omega$ depends mass $M_0$ and $m_0$ moving celestial bodies. And as energy depends on mass energy of a celestial body depends on dispersion $D\Omega$. And

$$D\Omega = \int_{-\infty}^{\infty} \omega^2 V(\omega) d\omega = \int_{-\infty}^{\infty} \omega^2 \left[ \frac{1}{\pi} \int_{0}^{\infty} y(t) \cos \omega t dt \right] d\omega$$

as from the Appendix 1 it is had

$$V(\omega) = \frac{1}{\pi} \int_{0}^{\infty} y(t) \cos \omega t dt .$$

From the received result follows, that

*the celestial body earns energy as a result of the moution.*

■ *N.A.Kozyrev's theoretical researches.*

In [6] on p. 197 of the section «The nature of stellar energy based on the analysis of observational data» N.A. Kozyrev writes: "Stars radiate in such a way as if they had been using the thermal and potential energy of their resources in accordance with the Helmholtz-Kelvin's mechanism. These resources, however, are rather limited. For instance, for Sun the life duration is only 30 million years which contradicts irrevocably the data of geology and cosmogony. It means that in reality the loss of energy does not cause the restructuring of the star. It gives rise to processes compensating these losses ... *and the star becomes a machine, generating energy*" (marked out by the author).

The conclusion by N.A. Kozyrev is confirmed by a result obtained using the model (3.3) ■

## 6. Correlation equations as equations of unperturbed Kepler's motion

It is known that the movement of a celestial body in accordance with equations (2.1) is a plane curve in a three dimensional space. Assuming that the choice of the coordinate system is in the hands of the researcher we can consider not a three dimensional systems of coordinates but a two-dimensional coordinate system x0y. Hence, we can and we will be considering only the first two equations in (2.1). Denote

$$\frac{\mu}{r^3} = \omega^2 \qquad (6.1)$$

in (2.1), and consider a solution of these two equations in the parametric form

$$x = A\sin\omega t, \quad y = B\cos\omega t. \tag{6.2}$$

The equation for the motion of a celestial body in the plane x0y can be written as

$$\frac{x^2}{A^2} + \frac{y^2}{B^2} = 1. \tag{6.3}$$

This is the equation of a circle if A = B, and the equation of an ellipse if A ≠ B. It is easily seen that second derivatives of functions (6.2) are related as (6.3), by virtue of

$$\frac{x''^2}{\omega^4 A^2} + \frac{y''^2}{\omega^4 B^2} = 1.$$

From (6.3) it follows that maximal values of the functions and of the second derivatives modulo are not equal to each other if the motion is ellipsoidal:

$$\max|x(t)| \neq \max|y(t)|, \quad \max|x''(t)| \neq \max|y''(t)|. \tag{6.4}$$

To be able to use this result we will consider identities of type (3.3) along axes x and y, respectively:

$$-\int_0^\infty \omega^2 V_x(\omega)\cos\omega\tau_x d\omega - a_x \int_0^\infty \omega V_x(\omega)\sin\omega\tau_x d\omega + b_x \int_0^\infty V_x(\omega)\cos\omega\tau_x d\omega \equiv 0, \quad (\tau > 0)$$

$$-\int_0^\infty \omega^2 V_y(\omega)\cos\omega\tau_y d\omega - a_y \int_0^\infty \omega V_y(\omega)\sin\omega\tau_y d\omega + b_y \int_0^\infty V_y(\omega)\cos\omega\tau_y d\omega \equiv 0. \quad (\tau > 0)$$

Assume that $\tau = 0$ and extending the limits of integration in that identities and, as mentioned above, we obtain:

$$D\Omega_x = b_x, \quad D\Omega_y = b_y,$$

Simultaneously, however, $D\Omega_x$ and $D\Omega_y$ are maximal modulo values of the second derivatives of functions $h_x(\tau)$ and $h_y(\tau)$, respectively. In para. 4 we already noted that equations (4.2) are the equations of the same type as the equations (2.1). Hence, second derivatives in (6.4) can be substituted with second derivatives of functions $h_x(\tau)$ and $h_y(\tau)$, i.e. with dispersions $D\Omega_x$ and $D\Omega_y$. But then we will have an inequality

$$D\Omega_x \neq D\Omega_y,$$

hence, follows the inequality

$$b_x \neq b_y. \tag{6.5}$$

For axes x and y we can write down equations of type (4.2) as

$$\frac{d^2 h_x}{d\tau_x^2} = -b_x h_x, \quad \frac{d^2 h_y}{d\tau_y^2} = -b_y h_y \tag{6.6}$$

or as

$$\frac{d^2 h_x}{d\tau_x^2} = -b_x h_x, \quad \frac{d^2 h_y}{d\tau_{\alpha y}^2} = -\alpha^2 b_y h_y, \tag{6.7}$$

taking into account that the substitution (3.5) can change the argument $\tau$, for instance, in the function $h_y(\tau)$, changing at the same time the value of the constant $b_y$, as already noted above. If the parameter $\alpha$ in (6.7) is chosen so that we have an equality

$$\omega_x^2 = b_x = \alpha^2 b_y = \omega_y^2,$$

then it is obvious that in (6.7) we will also have an equality

$$\tau_x = \tau_{\alpha y}.$$

Consequently, due to the inequality (6.5) in the equations (6.6)

$$\tau_x \neq \tau_y. \tag{6.8}$$

If in (6.6) we have $b_x = b_y$, i.e. the dispersions of random values along axes x and y are equal, then obviously we obtain an equation of type (6.3) for $A = B$, i.e. the circumference equation.

If, however, in (6.6) $b_x \neq b_y$, i.e. $\omega_x \neq \omega_y$, then we have the following solutions from these equations

$$h_x(\tau_x) = \cos\omega_x\tau_x, \qquad h_y(\tau_y) = \cos\omega_y\tau_y, \tag{6.9}$$

in which, unlike decisions (6.2), any constants are not present. From (6.9) these solutions also follows an equation of type (6.3) for $A \neq B$, i.e. the equation of an ellipse since by virtue of (5.1) we have an equation

$$\sin^2 \omega_x\tau_x + \cos^2 \omega_y\tau_y = 1,$$

though $\omega_x \neq \omega_y$, but $\omega_x\tau_x = \omega_y\tau_y$.

Consequently, equations of motion for any celestial body in the Solar system are written as (2.1) or (6.6), if the motion of the body follows a circle. In this case we have equalities

$$b_x = b_y, \qquad \tau_x = \tau_y,$$

i.e. systems of equations (2.1) and (6.6) are the same systems of equations.

If, however, the motion of a celestial body in the Solar system follows the elliptical trajectory, then we can correctly describe this motion only writing an equation of motion as (6.6), and in this case we have inequalities

$$b_x \neq b_y, \qquad \tau_x \neq \tau_y. \tag{6.10}$$

It immediately follows that the system (6.6) provides a fundamentally more accurate description of the motion of a celestial body compared to (2.1), because it includes the motion which is not covered by (2.1). Thus the inequality (6.5) in the equations (2.1) is essentially impossible, as frequency of rotation of a body on axes x and at is not identical. Hence, we shall not receive the equation (6.3), and it will not coincide any more with an observable trajectory of a celestial body.

## 7. Harmonic oscillation velocity is not constant

We will show now that in the case of elliptic motion the velocity of propagation of harmonic oscillation along axes x and y is not the same. Consider the oscillation velocity $v_x$ along the axis x, endowing the oscillation frequency $\nu_x$ also with indices:

$$v_x = \lambda \nu_x.$$

The wavelength does not have any index because it is known that the velocity of the wave of any length is the same. Multiply both sides of the velocity formula by $2\pi\tau_x$ and rewrite it as

$$2\pi\tau_x v_x = 2\pi\tau_x \lambda \nu_x = \lambda\omega_x\tau_x = C_1\lambda,$$

where

$$C_1 = \omega_x \tau_x.$$

Choice of an arbitrary constant value $C_1$, as noted in para.3, is done by the researcher, hence

$$v_x = \frac{\lambda C_1}{2\pi\tau_x} = C\frac{\lambda}{\tau_x}, \quad \left(C = \frac{C_1}{2\pi}\right). \tag{7.1}$$

The velocity of oscillation propagation along the y axis is specified by a formula similar to (7.1):

$$v_y = C\frac{\lambda}{\tau_y} \tag{7.2}$$

Note that the constant value C in formulas (7.1) and (7.2) is the same because

$$C_1 = \omega_x\tau_x = \omega_y\tau_y, \quad C = \frac{C_1}{2\pi}.$$

By virtue of (6.8) $\tau_x \neq \tau_y$, hence, the velocity of propagation of harmonic oscillation along axis x and y is not the same:

$$v_x \neq v_y.$$

Obviously, the resulting velocity in the plane x0y for $v_x \neq v_y$ (or, which is the same, for $b_x \neq b_y$) will differ from the resulting velocity which we obtain for $v_x = v_y$ (or $b_x = b_y$).

The difference in velocity should result and it does result in the rotation of the perihelion of planet Mercury observed in the Universe, different from the value calculated from the Newtonian gravity law. Since Mercury's rotation velocity is high and the orbit of the planet is a pronounced ellipse the above difference results in the divergence of 43 seconds of arc per century which needed an explanation. So, here is the explanation, or rather, the proof based on the theory of differential equations co-authored by I. Newton. We can conclude now that the <u>law of gravity enables us determining correctly the value of the coefficient in the equations of the unperturbed Keplerian motion for the case of the circular motion of a celestial body.</u>

## 8. One more property of the flexible (elastic) time

Consider once again (1.1) and remember that

$$\tau = t_1 - t_2,$$

where $\tau$ is a time interval between sections of the process $\xi(t)$ at arbitrary times $t_1$ and $t_2$. An expert on stationary processes would know that in obtaining the autocorrelation function of the process $\xi(t)$, (a $h(\tau)$ in (3.1) is an autocorrelation function of the process $\xi(t)$) the argument $\tau$ has to run necessarily through the values of not only $\tau < 0$ (the past time for us), but also $\tau > 0$ (the future time for us). For $\tau = 0$ (the current time for us) we have the dispersion of this process which is a constant value. If the model of the Universe (3.3) is adequate to its structure at research of the Universe we should observe simultaneously all three times:: present, past and future. Note that the first two properties, from the point of view of astrophysics, are not too unusual, unlike the third property - the future time. And, nevertheless, all three times were really observed in experiences.

■[9]**,** p. 76: «The observations aimed at measuring with a micrometer of the positions of the points in the sky which caused changes in electric conductivity of the resistor in the vicinity of these objects. It was shown that the changes originated from *three points in the sky* (marked out by the author):

1) position of the object *at the present moment* (marked out by the author),

2) position *in the past* (marked out by the author), with an accuracy to a refraction coinciding with the visible image, and

3) position *in the future* (marked out by the author), which the object will occupy when a signal from the Earth would have reached it at the speed of light»

[9]**,** p.78: "The observations described here were mainly done to find the empirical confirmation of this extremely responsible conclusion of a possibility to observe *future as already existing reality*" (marked out by the author)

[9], p. 89: «…everything which may happen *already exists in the future and persists in the past* (marked out by the author)».

[9], p. 92: «In this world the *future already exists and, hence, there is no surprise to be able to observe it now* (marked out by the author)». ■

## 9. Some other features of the model

### 1. Shift of spectral lines

The example given below is used in many monographs as another proof of Einstein's relativity theory. «It is commonly supposed that atomic oscillations can be treated as an ideal natural clock. The interval between the beginning and the end of oscillations of two identical atoms wherever measured should be the same. Assume that one atom is close to the surface of the Sun, while the other is on the surface of the Earth …». A conclusion usually made after analyzing this example runs as follows: «By virtue of the constant nature of the speed of light the oscillation of the atom of the Sun is somewhat slower than that of the Earth».

Consider the same issue within the context of the present model. The atom near the Sun is obviously heated more compared to the one on the surface of the Earth. Consequently, the dispersion (such question was considered in п.3) of the atom near the Sun is greater than that of the atom near the Earth. Substitution (3.5) in the case considered has the form

$$\omega_s = \frac{\omega_e}{\alpha},$$

where $\alpha > 1$. Hence, $\omega_s < \omega_e$, i.e. the conclusion is the same.

## 2. Deviation of the light beam

Already Newton admitted that light could have a weight, i.e. light has mass. If light has mass then irrespective of whose laws it obeys, Newton's or Einstein's, a light beam passing close to the Sun will have its trajectory deviation due to gravity. Calculations using Einstein's theory have come up with twice the results obtained by using Newton's theory. The experimental check-up has shown that calculations using Einstein's theory are much closer to real values than the ones using Newton's theory.

According to the model (3.3) additional deviation of the light beam close to the Sun is caused due to the greater heating of the beam since the beam equation parameters change. It is this phenomenon which causes a greater beam deviation closer to the Sun.

## 10 . Conclusions

1. Flexible (elastic) time does not depend on the space.

2. Universe is stationary and, consequently, Euclidean.

3. Flexible (elastic) time in the Universe can slow down, accelerate or even reverse.

4. The advance of Mercury's perihelion versus the predictions in accordance with the universal

    gravity law results from the inequality of the coefficients in the equations of the unperturbed

    Keplerian motion along axes x, y and z.

5. The velocity of propagation of harmonic oscillation in the Universe is not constant.

6. Long-range interaction, i.e. instantaneous communication between any two points of space
   in the Universe is possible.

7. Universe is a closed-loop stable-state information-energy system.

8. Universe revives the nature and acts as a barrier to the second law of thermodynamics where
stars are treated as machines which accumulate energy by moving.
9. Physics in the Universe is conceptually different from that of the Earth and, respectively, needs methods of investigation different from the ones which are used today. Specifically, the analysis of different phenomena using the Doppler effect, i.e. by the method of non stationary environment is not acceptable.

The model of the Universe is actually

I.Newton's specified, added and expanded model.

References

1. Feller W. An Int. to Prob. Theory and Its Appl., v. II. N-Y, London, Sydney, 1966.
2. Hovsepian F. A. Positive definite functions and the Universe structure. Proc. of the 4$^{rd}$ Inter. Conf. SICPRO'05 Institute of Control Sciences Publ. Moscow, 2005.
3. Hovsepian F. A. Solution of a linear homogeneous asymptotically stable differential equation as
a characteristic function of a random value. YI Inter. Social Congress « Globalization: the present and the future Russia ». Materials of performances, v.II, p. 464 – 465. RSSU Publ. Moscow, 2006.
4. Hovsepian F. A. Invalidity of the four-dimensional space-time concept. Proc. of the 7$^{rd}$ Inter. Conf. SICPRO'08 Institute of Control Sciences Publ. Moscow, 2008 (in print).
5. Kozyrev N.A. Selected works. Leningrad University Publishers, 1991.
6. Kozyrev N.A. Astronomical proof of the validity of Minkovski's four-dimensional geometry Минковского // Manifestation of cosmic factors on Earth and stars// Astrometry and celestial mechanics. M.-L., 1980, pp.85-93. (Problems in the studies of the Universe, issue 9).
7. Kozyrev N.A., Nasonov V.V. A new method of trigonometric parallaxes based on the measurement of the difference between the true and visible positions of a star // Astrometry and celestial mechanics. M.-L., 1978, pp.168-179. (Problems in the studies of the Universe, issue 7).
8. Kozyrev N.A., Nasonov V.V. Some properties of time revealed by astronomical observations// Manifestation of cosmic factors on Earth and stars// M.-L., 1980, pp.76-84. (Problems in the studies of the Universe, issue 9).
9. Khinchin A.Y. Korrelationstheorie des stationaren stochastischen Prozesse. Mathematische Annalen, v.109, 1934, pp. 604-615.
5. Korn G.A., Korn T.M. Mathematical Handbook for Scientists and Engineers. McGraw Hill Company, Inc. N-Y, Toronto, London, 1961.

If function h(τ) in (1.1) is characteristic function of some random value, in return Fourier transformation of this function

$$V(\omega) = \frac{1}{2\pi} \int_{-\infty}^{\infty} \exp(-i\omega\tau) h(\tau) d\tau$$

function $V(\omega) \geq 0$ for all $\omega$ and this condition – necessary and sufficient [1].

The specified property of function $V(\omega)$ is used in the proof of the theorem 1.

**Theorem 1**. Let $y_i(t)$ be a solution of an ordinary homogeneous asymptotically stable differential equation with constant real-valued coefficients, which is determined:

1) by any pair of complex-conjugated roots for i = 1,

2) by any pair of different real-valued roots for i = 2,

3) by one real-valued root of double multiplicity for i = 3

with initial conditions

$$\frac{dy_i(t)}{dt}\bigg|_{t=0} = 0, \; y_i(t)\big|_{t=0} = 1.$$

Then the function $h_i(\tau)$, consisting of $y_i(t)$ for $t \geq 0$ and $y_i(t)$ continued in the even manner to $t < 0$, is characteristic function for all i = 1, 2, 3.

*Proof.*

1. Let the solution $y_1(t)$ be determined by complex-conjugated numbers – $\chi \pm i\lambda$, i.e.

$$y_1(t) = (A\cos\lambda t + B\sin\lambda t)\exp(-\chi t). \quad (\chi > 0, \lambda > 0, t > 0)$$

For initial conditions we obtain:

$$y(t)\big|_{t=0} = A = 1, \qquad \frac{dy(t)}{dt}\bigg|_{t=0} = \frac{B}{A} = \frac{\chi}{\lambda}.$$

Hence, we rewrite $y_1(t)$ in view of the initial conditions as

$$y_1(t) = (\cos\lambda t + \frac{\chi}{\lambda}\sin\lambda t)\exp(-\chi t).$$

The $y_1(t)$ is a solution of the differential equation

$$\frac{d^2 y}{dt^2} + a\frac{dy}{dt} + by = 0, \quad (t \geq 0).$$

To obtain an even continuation of $y_1(t)$ to $t < 0$ and for this continuation to be a solution of the differential equation it is necessary to change the sign of the odd derivative in the above second order equation:

$$\frac{d^2y}{dt^2} - a\frac{dy}{dt} + by = 0. \quad (t \leq 0)$$

However, then by virtue of the Vieta's theorem the sign of the real-valued complex-conjugated root will also change, i.e. we will have $\chi < 0$.

For the first derivative in zero on the left and on the right to coincide it is needed along with the changed sign of $\chi$ to change the sign of $\lambda$ as well. Consequently, an even representation $y_1(t)$ should necessarily look as

$$y^*_1(t) = (\cos\lambda_1 t + \frac{\chi_1}{\lambda_1}\sin\lambda_1 t)\exp(-\chi_1 t). \quad (\chi_1 < 0, \lambda_1 < 0, t < 0)$$

Obtain an even function

$$h_1(t) = y_1(-t) = y_1^*(t) \text{ при } t \leq 0, \qquad h_1(t) = y_1(t) \text{ при } t \geq 0.$$

The Fourier transformation $V_1(i\omega)$ of the function $v_1(t)$ is equal to

$$V_1(i\omega) = \frac{1}{2\pi}\int_{-\infty}^{\infty} h_1(t)\exp(-i\omega t)dt = \frac{1}{2\pi}[\int_{-\infty}^{0} y_1^*(t)\exp(-i\omega t)dt + \int_0^{\infty} y_1(t)\exp(-i\omega t)dt] =$$

$$= \frac{1}{2\pi}[\int_0^{\infty} y_1^*(-s)\exp(i\omega s)ds + \int_0^{\infty} y_1(t)\exp(-i\omega t)dt].$$

Since according to our designations $y_1^*(-s) = y_1(s)$, the intervals in $V_1(i\omega)$ differ one from the other only by the sign of $\omega$, hence, their sum is a real-valued function, which permits writing the Fourier transformation of the function $v_1(t)$ as

$$V_1(\omega) = \frac{1}{2\pi}\int_{-\infty}^{\infty} h_1(t)\exp(-i\omega t)dt = \frac{1}{\pi}\int_0^{\infty} y_1(t)\cos\omega t\, dt.$$

Substituting in $V_1(\omega)$ the function $y_1(t)$ and calculating this intergral using (A3.1) and (A3.2) from Appendix 3, assuming that $\alpha = \chi$, $\beta = \lambda$, $\gamma = \omega$:

$$V_1(\omega) = \frac{1}{\pi}\int_0^{\infty}(\cos\lambda t + \frac{\chi}{\lambda}\sin\lambda t)\exp(-\chi t)\cos\omega t\, dt =$$

$$= \frac{1}{\pi}\{\frac{\chi(\chi^2 + \lambda^2 + \omega^2)}{[\chi^2 + (\lambda+\omega)^2][\chi^2 + (\lambda-\omega)^2]} + \frac{\chi}{\lambda}\frac{\lambda(\chi^2 + \lambda^2) - \lambda\omega^2}{[\chi^2 + (\lambda+\omega)^2][\chi^2 + (\lambda-\omega)^2]}\} =$$

$$= \frac{2\chi(\chi^2 + \lambda^2)}{\pi[\chi^2 + (\lambda+\omega)^2][\chi^2 + (\lambda-\omega)^2]} \geq 0$$

for all $-\infty < \omega < \infty$.

2. Let solution $y_2(t)$ be determined by different real-valued numbers $-\chi_1$ and $-\chi_2$, i.e.

$$y_2(t) = A\exp(-\chi_1 t) + B\exp(-\chi_2 t),$$

which assuming the initial conditions can be written as

$$y_2(t) = \frac{\chi_2}{(\chi_2 - \chi_1)}[\exp(-\chi_1 t) - \frac{\chi_1}{\chi_2}\exp(-\chi_2 t)].$$

The above used in obtaining an even function for $y_1(t)$ is also valid for obtaining an even function $y_2(t)$. Obtain this even function $h_2(t)$ as

$$h_2(t) = y_2(-t) = y_2^*(t) \text{ for } t \leq 0, \qquad h_2(t) = y_2(t) \text{ for } t \geq 0,$$

whose Fourier transformation can be presented as

$$V_2(\omega) = \frac{1}{\pi}\int_0^\infty y_2(t)\cos\omega t\, dt = \frac{1}{\pi}\int_0^\infty \frac{\chi_2}{(\chi_2 - \chi_1)}[\exp(-\chi_1 t) - \frac{\chi_1}{\chi_2}\exp(-\chi_2 t)]\cos\omega t\, dt.$$

Using the value (A3.1) from Appendix 3, assuming $\beta = 0$, $\gamma = \omega$, for the first integral $\alpha = \chi_1$, and for the second integral $\alpha = \chi_2$, we obtain

$$V_2(\omega) = \frac{1}{\pi}\frac{\chi_2}{(\chi_2-\chi_1)}\{\frac{\chi_1}{\chi_1^2+\omega^2} - \frac{\chi_1}{\chi_2}\frac{\chi_2}{\chi_2^2+\omega^2}\} = \frac{\chi_2}{\pi(\chi_2-\chi_1)}\frac{\chi_1(\chi_2^2+\omega^2)-\chi_1(\chi_1^2+\omega^2)}{\pi(\chi_1^2+\omega^2)(\chi_2^2+\omega^2)} =$$

$$= \frac{\chi_1\chi_2(\chi_1+\chi_2)}{\pi(\chi_1^2+\omega^2)(\chi_2^2+\omega^2)} \geq 0$$

for all $-\infty < \omega < \infty$.

3. Let solution $y_3(t)$ be determined by a real-valued-number $-\chi$ of double multiplicity, i.e.

$$y_3(t) = A\exp(-\chi t) + Bt\exp(-\chi t),$$

which assuming the initial conditions is given as

$$y_3(t) = \exp(-\chi t) + \chi t\exp(-\chi t).$$

In this case too we can obtain an even function $h_3(t)$ in accordance with the above rule. The two-sided Fourier transformation of the function $v_3(t)$ is equal to

$$V_3(i\omega) = \frac{1}{2\pi}\int_{-\infty}^\infty h_3(t)\exp(-i\omega t)dt = \frac{1}{\pi}\int_0^\infty y_3(t)\cos\omega t\, dt =$$

$$= \frac{1}{\pi}\int_0^\infty [\exp(-\chi t) + \chi t\exp(-\chi t)]\cos\omega t\, dt.$$

Utilizing (A3.1) from Appendix 3, assuming $\beta = 0$, $\gamma = \omega$, $\alpha = \chi$ for the first integral and (A3.4) from Appendix 3, we can write

$$V_3(\omega) = \frac{1}{\pi}\{\frac{\chi}{\chi^2+\omega^2} + \frac{\chi(\chi^2-\omega^2)}{(\chi^2+\omega^2)^2}\} = \frac{\chi(\chi^2+\omega^2)+\chi(\chi^2-\omega^2)}{\pi(\chi^2+\omega^2)^2} = \frac{2\chi^3}{\pi(\chi^2+\omega^2)^2} \geq 0$$

for all $-\infty < \omega < \infty$.

The Fourier transformation $V_i(\omega)$ of the function $h_i(\tau)$ (i = 1, 2, 3) for all types of roots set in the theorem 1 is a non-negative function. Consequently, $V_i(\omega)$ is the probability distribution density of a random value $\Omega_i$. Since $h_i(\tau)$ is a continuous function and

$$\int_{-\infty}^{\infty} |h_i(\tau)| d\tau < \infty,$$

then $h_i(\tau)$, as known, can be presented as a Fourier integral

$$h_i(\tau) = \int_{-\infty}^{\infty} \exp(i\omega\tau) V(\omega) d\omega = 2\int_{0}^{\infty} V_i(\omega) \cos\omega\tau \, d\omega.$$

Function $h_i(\tau)$ for all $i = 1, 2, 3$ is characteristic function by virtue of the Bochner's lemma [1].

This proves Theorem 1.

Appendix 2

**Theorem 2**. A characteristic function

$$h_i(\tau) = 2\int_{0}^{\infty} V_i(\omega) \cos\omega\tau \, d\omega \quad (i = 1, 2, 3; \; -\infty < \tau < \infty)$$

twice differentiated under the sign of integration with respect to $\tau$ as a parameter, i.e.

$$\frac{d^j h_i}{d\tau^j} = 2\frac{d^j}{d\tau^j}\int_{0}^{\infty} V_i(\omega) \cos\omega\tau \, d\omega = 2\int_{0}^{\infty} V_i(\omega) \frac{\partial^j \cos\omega\tau}{\partial\tau^j} d\omega, \quad (j = 1, 2).$$

*Proof.* Function $h_i(\tau)$ can be differentiated under the sign of integration with respect to $\tau$, if the improper integrals which result from the differentiation converge uniformly [10]. It is known that an improper integral

$$\int_{0}^{\infty} f(x, z) dx$$

converges uniformly on each set S of values z, if for any $z \in S$ and any x from the integration range the following inequality holds $|f(x, z)| \leq g(x)$, where $g(x)$ is the comparison function, whose integral converges:

$$\int_{0}^{\infty} g(x) dx < \infty.$$

According to the above feature the integral

$$\int_{0}^{\infty} \frac{\partial^2}{\partial\tau^2} V_i(\omega) \cos\omega\tau \, d\omega = -\int_{0}^{\infty} \omega^2 V_i(\omega) \cos\omega\tau \, d\omega$$

converges uniformly since $|\omega^2 V_i(\omega)\cos\omega\tau| \leq \omega^2 V_i(\omega)$, and

$$\int_{0}^{\infty} \omega^2 V_i(\omega) d\omega < \infty,$$

Which can be easily checked by direct integration. Now it is obvious that the following integral also converges uniformly

$$\int_{0}^{\infty} \frac{\partial}{\partial\tau} V_i(\omega) \cos\omega\tau \, d\omega = -\int_{0}^{\infty} \omega V_i(\omega) \sin\omega\tau \, d\omega,$$

since
$$|\omega V_i(\omega)\sin\omega\tau| \leq \omega V_i(\omega) \text{ при } \omega \geq 0,$$
and
$$\int_0^\infty \omega V_i(\omega)d\omega < \infty,$$

which can be easily checked by direct integration. From the convergence of these integrals it follows that the function $h_i(\tau)$ is twice differentiated with respect to $\tau$ as a parameter for all i = 1, 2, 3 under the sign of integration.

This proves Theorem 2.

<div align="right">Appendix 3.</div>

<div align="center">Auxiliary propositions</div>

*Proposition* 1. If $\alpha > 0$, then

$$\int_0^\infty \exp(-\alpha t)\cos\beta t\cos\gamma t\, dt = \frac{\alpha[\alpha^2+\beta^2+\gamma^2]}{[\alpha^2+(\beta+\gamma)^2][\alpha^2+(\beta-\gamma)^2]}, \tag{A3.1}$$

$$\int_0^\infty \exp(-\alpha t)\sin\beta t\cos\gamma t\, dt = \frac{[\beta(\alpha^2+\beta^2)-\beta\gamma^2]}{[\alpha^2+(\beta+\gamma)^2][\alpha^2+(\beta-\gamma)^2]}. \tag{A3.2}$$

*Proof of Proposition* 1. Calculate the integral

$$\int_0^\infty \exp[-\alpha t + i(\beta\pm\gamma)t]\, dt = \frac{1}{-\alpha+i(\beta\pm\gamma)}\int_0^\infty d\{\exp[-\alpha t + i(\beta\pm\gamma)t]\} =$$

$$= -\frac{1}{-\alpha+i(\beta\pm\omega)} = \frac{\alpha+i(\beta\pm\gamma)}{\alpha^2+(\beta\pm\gamma)^2}. \tag{A3.3}$$

Note that integrals in the proposition 1 can be easily calculated with (A3.3). The first integral

$$\int_0^\infty \exp(-\alpha t)\cos\beta t\cos\gamma t\, dt = \frac{1}{2}\text{Re}\{\int_0^\infty [\exp(-\alpha t + i\beta t)][\exp(i\gamma t)+\exp(-i\gamma t)]dt\} =$$

$$= \frac{1}{2}\{\frac{\alpha}{\alpha^2+(\beta+\gamma)^2}+\frac{\alpha}{\alpha^2+(\beta-\gamma)^2}\} = \frac{\alpha[\alpha^2+\beta^2+\gamma^2]}{[\alpha^2+(\beta+\gamma)^2][\alpha^2+(\beta-\gamma)^2]}.$$

The second integral

$$\int_0^\infty \exp(-\alpha t)\sin\beta t\cos\gamma t\, dt = \frac{1}{2}\text{Im}\{\int_0^\infty [\exp(-\alpha t + i\beta t)][\exp(i\gamma t)+\exp(-i\gamma t)]dt\} =$$

$$= \frac{1}{2} \left\{ \frac{\beta+\gamma}{\alpha^2+(\beta+\gamma)^2} + \frac{\beta-\gamma}{\alpha^2+(\beta-\gamma)^2} \right\} =$$

$$= \frac{1}{2} \left\{ \frac{(\beta+\gamma)[\alpha^2+(\beta-\gamma)^2]}{[\alpha^2+(\beta+\gamma)^2][\alpha^2+(\beta-\gamma)^2]} + \frac{(\beta-\gamma)[\alpha^2+(\beta+\gamma)^2]}{[\alpha^2+(\beta+\gamma)^2][\alpha^2+(\beta-\gamma)^2]} \right\} =$$

$$= \frac{(\beta+\gamma)(\alpha^2+\beta^2-2\beta\gamma+\gamma^2)+(\beta-\gamma)(\alpha^2+\beta^2+2\beta\gamma+\gamma^2)}{[\alpha^2+(\beta+\gamma)^2][\alpha_2+(\beta-\gamma)^2]} =$$

$$= \frac{\beta(\alpha^2+\beta^2)-\beta\gamma^2}{[\alpha^2+(\beta+\gamma)^2][\alpha^2+(\beta-\gamma)^2]}.$$

Proposition 1 is proved.

*Proposition* 2. If $\alpha > 0$, then

$$\int_0^\infty t\exp(-\alpha t)\cos\omega t\, dt = \frac{\alpha^2-\omega^2}{(\alpha^2+\omega^2)^2}. \qquad (A3.4)$$

*Proof of Proposition* 2. This integral can be easily calculated if note that

$$\int_0^\infty t\exp(-\alpha t)\cos\omega t\, dt = \int_0^\infty \frac{\partial}{\partial \omega}[\exp(-\alpha t)\sin\omega t]dt.$$

Then using the value of the second integral from Proposition 1, we can use the theorem of differentiation with respect to a parameter [10]:

$$\int_0^\infty \frac{\partial}{\partial \omega}[\exp(-\alpha t)\sin\omega t]dt = \frac{d}{d\omega}\int_0^\infty \{\exp(-\alpha t)\sin\omega t\, dt\} = \frac{d}{d\omega}\frac{\omega}{\alpha^2+\omega^2} =$$

$$= \frac{\alpha^2+\omega^2-2\omega^2}{(\alpha^2+\omega^2)^2} = \frac{\alpha^2-\omega^2}{(\alpha^2+\omega^2)^2}.$$

This proves Proposition 2.